\documentclass[prl,showpacs,aps,twocolumn]{revtex4}
\usepackage{graphicx}
\usepackage{epstopdf}
\usepackage{dcolumn}
\usepackage{bm}
\usepackage{color}
\usepackage{ulem}
\usepackage{amsfonts,amsthm}
\usepackage{amsmath}
\usepackage{amssymb}
\begin{document}
\newcommand{\bR}{\mbox{\boldmath $R$}}
\newcommand{\tr}[1]{\textcolor{red}{#1}}
\newcommand{\trs}[1]{\textcolor{red}{\sout{#1}}}
\newcommand{\tb}[1]{\textcolor{blue}{#1}}
\newcommand{\tbs}[1]{\textcolor{blue}{\sout{#1}}}
\newcommand{\Ha}{\mathcal{H}}
\newcommand{\mh}{\mathsf{h}}
\newcommand{\mA}{\mathsf{A}}
\newcommand{\mB}{\mathsf{B}}
\newcommand{\mC}{\mathsf{C}}
\newcommand{\mS}{\mathsf{S}}
\newcommand{\mU}{\mathsf{U}}
\newcommand{\mX}{\mathsf{X}}
\newcommand{\sP}{\mathcal{P}}
\newcommand{\sL}{\mathcal{L}}
\newcommand{\sO}{\mathcal{O}}
\newcommand{\la}{\langle}
\newcommand{\ra}{\rangle}
\newcommand{\ga}{\alpha}
\newcommand{\gb}{\beta}
\newcommand{\gc}{\gamma}
\newcommand{\gs}{\sigma}
\newcommand{\vk}{{\bm{k}}}
\newcommand{\vq}{{\bm{q}}}
\newcommand{\vR}{{\bm{R}}}
\newcommand{\vQ}{{\bm{Q}}}
\newcommand{\vga}{{\bm{\alpha}}}
\newcommand{\vgc}{{\bm{\gamma}}}
\arraycolsep=0.0em
\newcommand{\Ns}{N_{\text{s}}}

\title{
{\it Ab initio} Evidence for Strong Correlation Associated with Mott Proximity \\
in Iron-based Superconductors
}

\author{Takahiro Misawa,
Kazuma Nakamura, and
Masatoshi Imada}

\affiliation{%
Department of Applied Physics, University of Tokyo, and JST CREST,
7-3-1 Hongo, Bunkyo-ku, Tokyo, 113-8656, Japan}

\date{\today}
\begin{abstract}
We predict that iron-based superconductors discovered near $d^6$ 
configuration (5 Fe 3$d$ orbitals filled by 6 electrons) is located on the 
foot of an unexpectedly large dome of correlated electron matter centered at the 
Mott insulator at $d^5$ (namely, half filling). This is based on 
the many-variable variational Monte-Carlo results for $ab$ $initio$ 
low-energy models derived by the downfolding. The $d^5$ Mott 
proximity extends to subsequent emergence of incoherent metals, orbital 
differentiations due to the Mott physics and Hund's-rule coupling, followed by 
antiferromagnetic quantum criticality, in quantitative accordance with available experiments. 
\end{abstract}
%

\maketitle

{\it Introduction.---}
Iron-based superconductors discovered in 2008~\cite{Kamihara_LaFeAsO} soon proved to encompass different families of pnictides such as LaFeAsO and BaFe$_2$As$_2$ and chalcogenides such as FeSe and FeTe, where the superconductivity was discovered mostly under electron or hole doping as in the 
cuprate superconductors~\cite{Bednorz}. 
The backbone lattices are commonly built from a stacking of iron square-lattice layers. 
The band structures consisting mainly of the Fe 3$d$ orbitals located near the Fermi level are also similar among the families\cite{Miyake}.
However, physical properties strongly depend on the families and have a diversity in magnetic, transport and superconducting properties\cite{HosonoReview}. 

The antiferromagnetic (AF) order is, in most cases, found close to the superconductivity, 
while the ordered magnetic moment has a diversity from zero up to $\ge 2\mu_{\rm B}$\cite{Mook_Nature2008,Qureshi,Huang,Li_PRB2009} whose origin 
is not entirely clear. Another diversity is found in the coherence of the metallic carrier. 
The incoherent (``bad metallic") conduction with small 
Drude weight\cite{Boris,Yang_Timusk,Qazilbash}, enhanced mass in 
deHaas-van Alphen measurement\cite{Terashima}, strongly renormalized quasiparticle 
peak in ARPES\cite{Tamai}, unconventional AF fluctuations 
\cite{NMR} and a typical 
Mott-Hubbard splitting (
emergence of the lower Hubbard band)\cite{Aichhorn} have been claimed in several compounds, while the 
correlation effects are less clear in other cases\cite{Malaeb}. Such a rich diversity is 
characteristic in contrast to the cuprate superconductors, in which strong correlation 
effects are common and universal. In this letter, we first show that the diversity 
emerges indeed from the variation in the electron correlation.  

The strength of electron correlation in the present materials 
is presently under hot debate
partly because of this diversity.
Firm and accurate calculations with full account of not only dynamical but also spatial fluctuation effects are desired. 
However, {\it ab initio} analyses 
with full account of spatial correlation effects require demanding calculations and to our knowledge are very few.

In this letter, we employ an {\it ab initio} method\cite{ImadaMiyake} by a hybrid combination of density functional theory with accurate solvers for the downfolded effective low-energy model. Here, we employ the multi-variable variational Monte Carlo (mVMC) method\cite{Tahara,Misawa} for the solver. See Supplemental Materials S.1 and S.3 for details of the whole methods. The method enables us to examine strong correlation effects in an {\it ab initio} way fully with dynamical as well as spatial fluctuation effects. We find that the real iron-based  superconductors are on the foot of a large-dome structure centered around the $d^5$ Mott insulator, whose proximity effects generate various correlation phenomena of the real superconductors. This sheds new light on the understanding of the electron correlation effects in the iron-based materials.   
 
In general, to vary and control electronic properties in correlated electron systems, two important routes are known\cite{ImadaRMP}. One is the bandwidth (or equivalently effective-Coulomb-interaction) control and the other is the filling control. The former directly controls the ratio between the kinetic and interaction energies of electrons, while the latter tunes the distance from the ``commensurate filling" (a simple fractional number of band filling enhances the electron correlation as in the Mott insulator). We show that both play key roles in the iron-based superconductors.

{\it Model.---}
The derived {\it ab initio} parameters for the present effective models by the downfolding procedure\cite{Miyake,Nakamura_2D} of the Fe $3d$ orbitals consist of the transfer integrals $t_{i,j,\nu,\mu}$ of an electron between the orbital $\nu$ on the site $i$, and $\mu$ on $j$, together with the orbital-dependent onsite direct-Coulomb ($U_{\nu,\mu}$) and the exchange ($J_{\nu,\mu}$) interactions.   The ratio of the effective interactions (so called $U$ representing the orbital average of $U_{\nu,\mu}$) to the bandwidth (or averaged transfer $t$) substantially (approximately 50\%) increases in the order from LaFePO, LaFeAsO, BaFe$_2$As$_2$, FeTe to FeSe\cite{Miyake}. This is physically well understood by the increasing distance $h$ between an iron layer and the neighboring pnictogen/chalcogen atoms, which alters the chemical bonding of the Fe 3$d$ Wannier orbitals with the pnictogen/chalcogen $p$ orbitals from more covalent to more ionic in this order. This makes the Fe 3$d$ Wannier orbitals more localized and the screening by pnictogen/chalcogen $p$ and other orbitals less effective. Therefore, the family dependence offers a good example of the interaction control.  

The interaction $U_{\mu,\nu}$ strongly depends on the orbitals $\mu,\nu$, and compounds. Nevertheless, all the derived {\it ab initio} models are reasonably well reproduced from that of a particular compound, say LaFeAsO, by a single parameter $\lambda$ which scales all the interactions, $U_{\nu,\mu}$ and $J_{\nu,\mu}$ uniformly (see Supplemental Materials S.2). In fact, $\lambda$ is a measure of the electron correlation.

\begin{figure}[htp!]
	\begin{center}
		\includegraphics[width=7.2cm,clip]{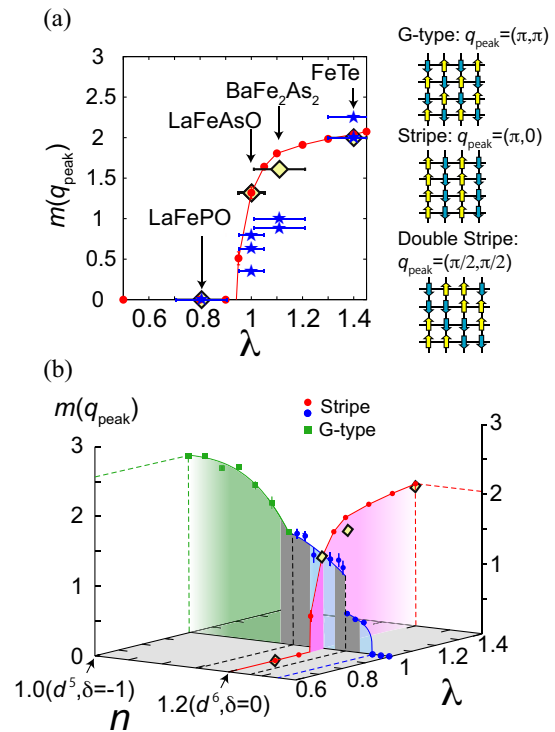}   
	\end{center}
\caption{(color).~(a)~Magnetic ordered moment $m(q_{\rm peak})$ at the Bragg wavenumber $q_{\rm peak}$ (for the pattern, see right) calculated by mVMC (red circles). The result at $\lambda$=1 represents that of the {\it ab initio} model for LaFeAsO while the $\lambda$ dependence illustrates the results of the models obtained by uniformly scaling the interaction strengths by $\lambda$. The black-framed yellow diamonds indicate the results of the {\it ab initio} models, where they are plotted at the corresponding $\lambda$. The plotted results are those in the thermodynamic limit after the size extrapolation. Experimentally observed values\cite{Mook_Nature2008,Qureshi,Huang,Li_PRB2009} are also plotted at corresponding $\lambda$ by asterisks, where FeTe shows double-stripe and others show stripe orders in agreement with our {\it ab initio} results. (b)~Magnetic ordered moment in the plane of $\lambda$ and doping concentration $\delta$. The data are plotted in the cross-sections for $\lambda$=1 as well as for $\delta$=0. It shows a peak at $d^5$ ($\delta$=$-$1.0) and decreases monotonically over $d^6$ ($\delta$=0), which forms a large half-dome structure peaked at $d^5$. The mVMC results of the {\it ab initio} models are shown by black-framed yellow diamonds. The green and blue shaded regions represent the G-type and stripe-type AF orders, respectively, as is illustrated in the right panel of Fig.\ref{Fig.1}~(a). There exist two first-order transitions (black dashed lines), one indicated by the jumps in the ordered moment around $\delta$$\sim$0.17 and the other at the transition between the G-type and stripe around $\delta$$\sim$$-$0.22, which signals large charge fluctuations under phase-separation effects. In the present short-ranged-interaction model, the phase separation indeed occurs in the gray shaded regions.
}
\label{Fig.1}
\end{figure}%

{\it Result.---}
In Fig.\ref{Fig.1}~(a), we show the ordered magnetic moment of the AF order $m(q_{\rm peak})$ (red circles) as a function of $\lambda$, calculated by mVMC with the extrapolation to the thermodynamic limit. 
Here, $\lambda=1$ represents the {\it ab initio} model for LaFeAsO, while by uniformly scaling all the interaction strength by $\lambda$, the interaction control is monitored. We also plot the mVMC result of the real {\it ab initio} models for LaFePO, BaFe$_2$As$_2$, and FeTe, as balck-framed yellow diamonds, at corresponding $\lambda$ values (see Supplemental Materials S.2)\cite{Misawa}. 
The result for the {\it ab initio} models of FeTe shows the double-stripe order degenerate with the simple stripe with a nearly equal ordered moment. 
All the other results shown here indicate the stripe-ordered ground states. The agreement with the experimental results with the correct AF quantum critical point proves the accuracy of the present scheme. Similar quantitative agreements with the experiments have been suggested in earlier studies\cite{Misawa,Yin}.  

The uncertainties in these $\lambda$-scaling plots for results of the experiments (blue asterisks) and the {\it ab initio} models are indicated by horizontal error bars. Here, the uncertainty arises from the fact that two independent {\it ab initio}-model derivations give a slight difference ($\sim 5\%$)~\cite{Miyake}, together with the error caused by details of orbital variations ignored in the uniform $\lambda$ scaling ($\sim 5\%$).

Quantum critical point of the AF transition appears at around $(\lambda \sim 0.95)$.The overall agreement among 
the $\lambda$-scaled models, {\it ab initio} models and the experiments show that the material dependence of the magnetism is well described by the variation of the correlation strength represented by the single parameter $\lambda$.

In addition to the interaction control, they may be substituted by other elements so that carriers are doped by holes or electrons (namely the filling control) as compared to the mother materials at $d^6$. This simultaneous possibility of the interaction and filling controls makes the iron-based superconductors valuable and unique in comparison to 
the cuprates (mainly, filling control only) or organic conductors (mainly, bandwidth control only)\cite{Kanoda}.  Furthermore, nearly degenerate five 3$d$ orbitals enable us to examine multi-orbital effects, that are absent in the cuprates and the organic conductors.  

In the plane of $\lambda$ as well as the doping concentration $\delta$, the magnetic ordered moment $m(q_{\rm peak})$ is plotted in Fig. \ref{Fig.1}~(b). Here, as in $\lambda$, $\delta$ monitors the doping effect, if the real material could be purely doped without changing other parameters such as the transfers. By this monitoring, we can elucidate physics of the filling control. As we clarify later, we caution that it does not simply mean the substitution effect of the real material, because the real substitution may change other parameters. Although the region near the $d^5$ configuration has not been experimentally realized so far, our monitoring of the filling control predicts that the iron-based superconductors around $\delta$=0 ($d^6$) are located in the foot of a large dome centered at $\delta$=$-$1.0 ($d^5$).  
It was formerly believed that metallic ``valleys" at non-integer fillings intervene the insulators formed at each integer filling (see for instance Fig.65 of Ref.~\cite{ImadaRMP}). In marked contrast, the monotonic decrease in $m(q_{\rm peak})$ from $d^5$ over $d^6$ supports that the electron correlation of the iron-based superconductors around $d^6$ emerges just as a proximity effect of the $d^5$ Mott insulator without a good metallic region between $d^5$ and $d^6$ for the retained interaction strength. 

\begin{figure}[htp!]
	\begin{center}
	 \includegraphics[width=8.5cm,clip]{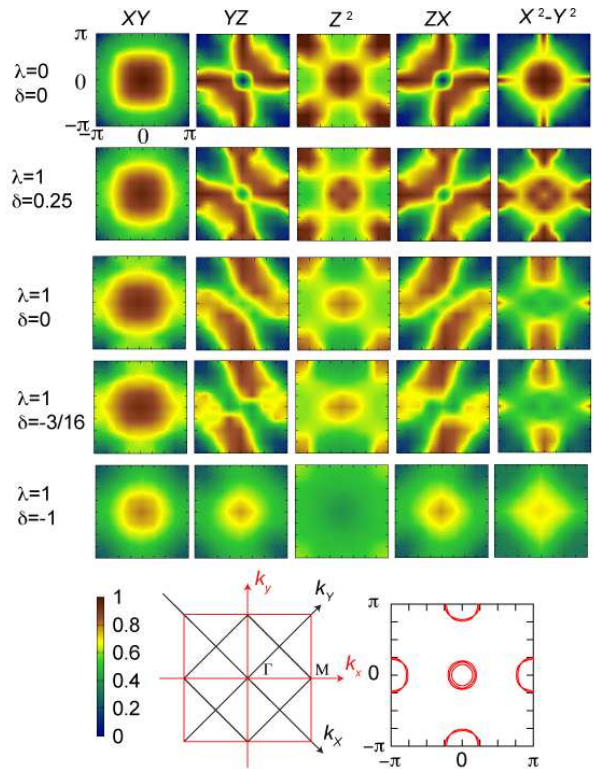}
	\end{center}
\caption{
(color).~Filling dependence of orbital-resolved momentum distribution $n_{\nu}(k)\equiv \langle c^{\dagger}_{\nu,k}c_{\nu,k}\rangle$ plotted for the ``unfolded" Brillouin zone (BZ), where the creation and annihilation operators of an electron at the orbital $\nu$ and the wavenumber $k$ are denoted by $c_{\nu,k}^{\dagger}$ and $c_{\nu,k}$, respectively. In the bottom left, the unfolded BZ (with the coordinates $k_x$ and $k_y$) and folded BZ (with $k_X$ and $k_Y$) are depicted by the red and black lines, respectively. In the bottom right, the Fermi surfaces of the LDA band structures are shown by the red curve, which can be identified in the corresponding sharp changes in $n_{\nu}(k)$ for $\nu=d_{YZ/ZX}$ and $d_{X^2-Y^2}$ on the top panels. Note that the orbitals are represented according to the folded BZ. Fade out of sharp boundaries in $n_{\nu}(k)$, representing the Fermi pockets, are seen especially for $d_{YZ/ZX}$ and $d_{X^2-Y^2}$ together with $d_{Z^2}$, which signals strong renormalizations of quasiparticles with bad metallic behavior when $\delta$ decreases progressively to negative. 
}
\label{Fig.2}
\end{figure}

Indeed the orbital-resolved momentum distribution $n_{\nu}(k)$ in Fig.\ref{Fig.2} shows that the clear jump manifested at the Fermi surface of good metals becomes less and less clear with decreasing electron concentration  from $d^6$ ($\delta$=0) toward $d^5$ ($\delta$=$-$1). The reason for the strong incoherence toward $d^5$ is that all the orbitals become nearly half filling at $d^5$. Such a strong commensurability leads to a solid Mott insulator.
The large dome structure is a consequence of an overwhelming strong proximity of the $d^5$ (half-filled) Mott insulator, which blankets the $d^6$ commensurability. 
Furthermore, signatures of several first-order transitions seen in Fig.\ref{Fig.1}~(b) in the filling control signal large charge fluctuations with tendency toward the phase separation and incoherence. 

Another important aspect found in the doping dependence is the switching in the magnetic order; the transition 
from  the stripe to G-type orders at $\delta \sim -0.22$ [Fig.1~(b)], which is explained by the geometrical frustration effects: The ratio of the diagonal next-nearest ($t'$) to the nearest-neighbor($t$) transfers, i.e., $t'/t$ measures the frustration effect. In fact, $d_{YZ}/d_{ZX}$ and $d_{X^2-Y^2}$ orbitals have $t'/t\sim 1.0$, while for $d_{XY}$ and $d_{Z^2}$ orbitals $t'/t\sim 0.1$~\cite{Miyake}, where the orbitals are labeled in the representation of the folded Brillouin zone~\cite{Miyake}. 
 Around $d^6$, the frustrated $d_{X^2-Y^2}$ is dominant because of the differentiated pinning to half filling, while near $d^5$,  all the orbitals are pinned to half filling and $d_{XY}$ and $d_{Z^2}$ become the more important players because of the larger effective interactions, which favors the G-type order. Nearby G-type order is suggestive of proposed nodes in the pairing symmetry of heavily hole doped KFe$_2$As$_2$\cite{KFe2As2-Fukazawa,KFe2As2-Hashimoto}. In this connection, the double-stripe type order at $(\pi/2,\pi/2)$ experimentally observed for FeTe is also accounted for by the involvement of the less frustrated $d_{Z^2}$ orbital. 

\begin{figure}[htp!]
\begin{center}
\includegraphics[width=8.5cm,clip]{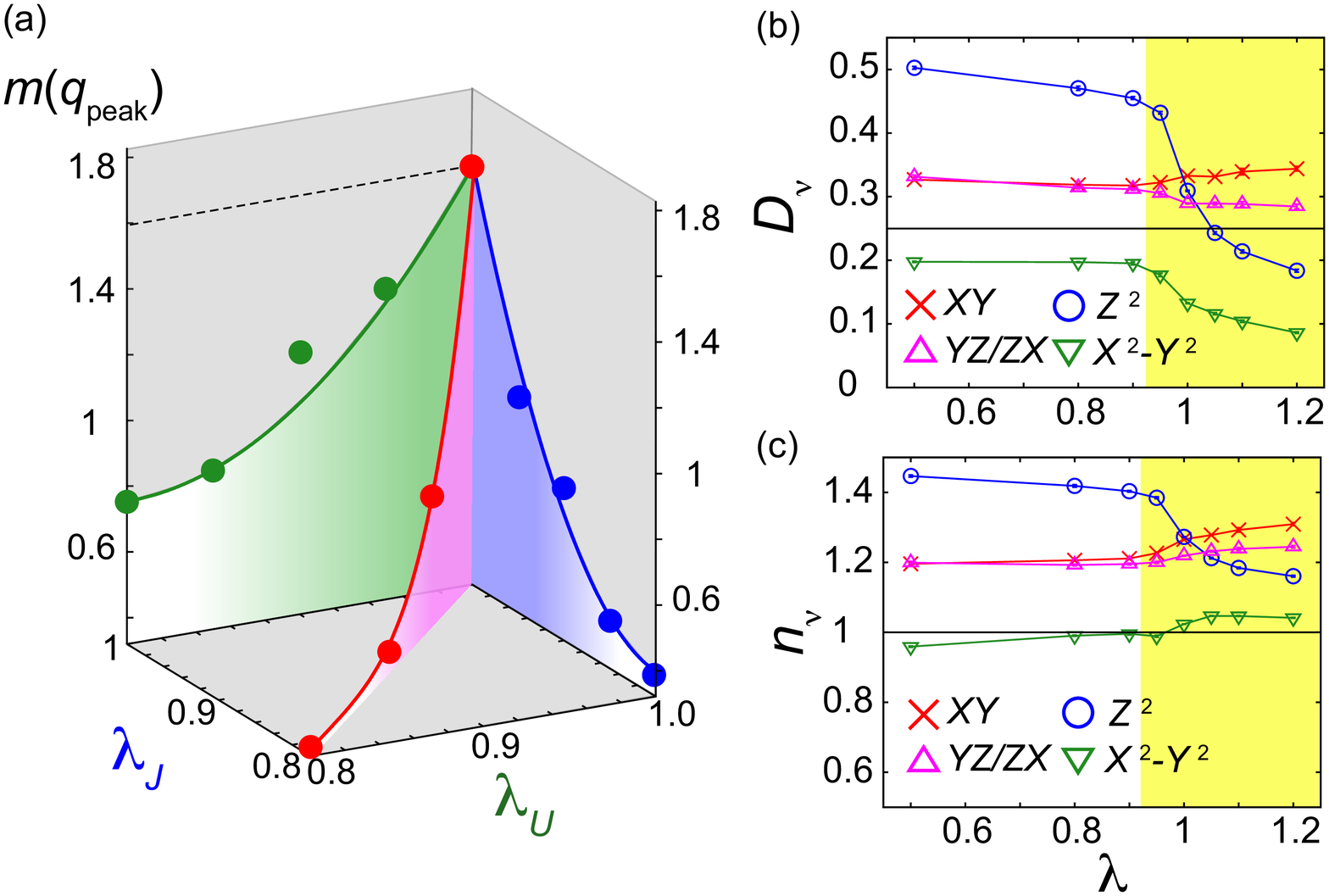}
\end{center}
\caption{(color online).~(a)~Magnetic ordered moments in the plane of $\lambda_U$ and $\lambda_J$ for system size $N_{\rm s} = 6\times 6$, which is expected to be close to the thermodynamic limit. Here, $\lambda_{J}$ ($\lambda_{U}$) scales exchange (Coulomb) interactions from the {\it ab initio} model for LaFeAsO at $\lambda_U=\lambda_J=1$. The sharp change in the ordered magnetic moment is triggered by a synergy effect of the direct Coulomb ($\lambda_U$) and the exchange ($\lambda_J$) interactions and the two interactions comparably contribute. (b)~Orbital-resolved double occupation $D_{\nu}$ defined by probability of simultaneous occupation of up and down spin electrons on the $\nu$ orbital at the same site. (c)~Orbital occupation $n_{\nu}$ defined by the averaged density of electrons on the orbital $\nu$ as a function of $\lambda$ at filling $\delta=0 (d^6$). Antiferromagnetic phase is represented as shaded region. System sizes $8\times 8$ for (b) and (c) are sufficient to well represent these quantities in the thermodynamic limit. Near the realistic parameter $\lambda=1$, strong correlation of the $d_{X^2-Y^2}$ orbital synergetically starts involving the $d_{Z^2}$ orbital through the Hund's rule coupling, which drives the $d_{Z^2}$ into the state close to half filling $n=1$ with reduced $D$.}
     \label{Fig.3}
  \end{figure}

To further understand the proximity effect of $d^5$ Mott physics, we monitor 
the 
effects of Coulomb ($U$) and exchange ($J$) interactions separately 
by the scaling parameters $\lambda_U$ and $\lambda_J$, respectively.
Figure \ref{Fig.3} reveals that  
$U$ and $J$ comparably contribute at $d^6$
in enhancing the ordered magnetic moment. 
The Hund's rule coupling $J$ 
retains the metallic but AF order even far away from half filling ($d^5$) contrary to the quick collapse of the AF order in the cuprates.
The present result conforms with a suggestion of an importance of $J$ away from half filling~\cite{Medici}.
We see below that $J$ between $d_{X^2-Y^2}$ and $d_{Z^2}$ orbitals is crucial in the iron families.

Around $d^6$, large spin and orbital fluctuations are prominent with the nearby magnetic quantum critical point. 
Figures \ref{Fig.3} (b) and (c) illustrate that the $d_{X^2-Y^2}$ orbital is under the strongest correlation effects indicated by the smallest double occupation [panel~(b)] and a remarkable pinning around half filling [(c)] even for relatively small $\lambda$, though the effective Coulomb interaction on the $d_{X^2-Y^2}$ is the smallest even at $\lambda=1$~\cite{Nakamura_2D}, i.e., onsite intraorbital interaction for the $d_{X^2-Y^2}$ orbital is 1.68 eV in comparison to 2.24-2.75 eV for other intraorbital values.~(See Table I in Supplemental Materials.) This puzzle is solved by the largest bare density of states at the Fermi level (see Fig.3 of Ref.~\cite{Miyake}), which sensitively allows the electron correlation through the electron-hole polarizations. When $\lambda$ increases around $\lambda=1$ (realistic for LaFeAsO), the $d_{Z^2}$ orbital quickly follows up the $d_{X^2-Y^2}$, because the $d_{X^2-Y^2}$  orbital has the largest Hund's rule coupling $J$ with $d_{Z^2}$ (0.43 eV in comparison to 0.23 eV with $d_{XY}$ and 0.35 eV with $d_{YZ/ZX}$). The moment at the $d_{Z^2}$ orbital grows, dragged by $d_{ X^2-Y^2}$ through $J$. We confirmed that such orbital differentiations occur commonly in the {\it ab initio} models for other families~\cite{Aichhorn}.The strong crossover from low to high magnetic moments by changing from LaFeAsO, BaFe$_2$As$_2$ through FeTe is accounted by this interplay of $d_{X^2-Y^2}$  and $d_{Z^2}$ orbitals.

One might suspect that the disappearance of the magnetic order in hole doped Ba$_{1-x}$K$_x$Fe$_2$As$_2$ at $x\sim 0.4$\cite{BFA} appears to contradict the present result. However, K doping does not work as a simple filling control but yields more complicated effects, because the Fe square lattice shrinks upon the K doping. This simultaneously enhances the bandwidth (nearly 10\%) and works also as the bandwidth control driving into weak correlation regime. At $\delta=-1$ and $\lambda=1$, we obtained the $d^5$ Mott insulator with a large Mott gap ($\sim 2.2$eV) with a high ordered moment ($\sim$ 2.5$\mu_{\rm B}$). It implies a connection with a G-type AF insulator in isostructural but $d^5$ compound LaMnPO\cite{Simonson}.

{\it Summary.---}
We have shown that the underlying proximity effects of the $d^{5}$ Mott insulator governs the electronic structures of iron-based superconductors, which are located around $d^{6}$ filling. When the Mott proximity is weakened, strong orbital and spin fluctuations take place before the verge to weakly correlated metals.  This is indeed the high-$T_c$ superconducting region around the $d^6$ filling.  It shares a common character with the cuprates, while a new aspect here is an involvement of the orbital fluctuation and differentiation under the big umbrella of the $d^5$ Mott insulator. 
Although $d^{6}$ is commensurate filling, the Dirac nodes in the 
band dispersions maintain the metal for unexpectedly wide region~\cite{Misawa,Ran}, while the 
interactions develop the antiferromagnetism.
The AF quantum critical points accompany the strong 
crossover of the orbital polarization as well and the 
superconducting mechanism has to be clarified under this circumstance. We propose to focus 
particularly on the role of the $d_{X^2-Y^2}$ and $d_{Z^2}$ orbitals 
as the leading players, in contrast to the $d_{YZ}/d_{ZX}$ orbitals in the nesting picture. 
In the present results, the $d^5$ dominance washes away the subtlety of the nesting. 
Experimental clarification of the large dome structure in the 
full bandwidth and filling controls in the range from $d^5$ to $d^6$ is 
highly desired to clarify the whole perspective of the electron correlation effect in the iron-based superconductors.

The authors are indebted to Takashi Miyake for valuable discussions and providing us
with his band structure data. They also thank Daisuke Tahara and Satoshi Morita for providing us with efficient mVMC codes. 
They are grateful to Toshiyuki Imamura for the usage of his matrix diagonalization code.
This work is financially supported by MEXT HPCI Strategic Programs for 
Innovative Research (SPIRE) and Computational Materials Science Initiative (CMSI).
The authors thank Laurence Livermore National Laboratory for generous offer of the supercomputer facilities.
Numerical calculation was partly carried out at the Supercomputer Center, Institute for Solid State Physics, Univ. of Tokyo. 
This work was also supported by Grant-in-Aid for 
Scientific Research (No. 22104010, 22340090, 22740215 and 23740261) from MEXT, Japan.

\clearpage
\noindent
{\Large Supplemental Materials}

\renewcommand{\theequation}{S.\arabic{equation}}
\setcounter{equation}{0}
\renewcommand{\tablename}{Table S}

\noindent
\section{ S.1 DERIVATION OF EFFECTIVE LOW-ENERGY MODELS 
BY AB INITIO PROCEDURE
}

To understand the normal state of iron-based superconductors, 
we employ the Multi-energy-scale {\it Ab initio} scheme for Correlated Electrons (MACE), which is a 
hybrid method of {\it ab initio} calculations by combining density functional theory with an 
accurate low-energy solver\cite{S_ImadaMiyake}. We first compute the global electronic structure by 
the local density approximation and 
eliminate
the bands far away from the Fermi level by a downfolding 
procedure after renormalizing
the effects of these eliminated bands.
The renormalization consists of 1) 
the construction of the Wannier orbitals for the target bands and the derivation of the tight-binding Hamiltonian 
in the Wannier basis followed by 2) the computation of the effective interaction for the target Wannier orbitals screened by the eliminated bands. 
This partial screening is computed by following the standard constrained RPA method\cite{S_ImadaMiyake}. 
After deriving the effective low-energy model,
3) it is solved by solvers such as the mult-variable 
variational Monte Carlo method~\cite{S_Misawa,S_Tahara} whose framework is outlined in S.3.
This hybrid method has been applied to various materials that exhibit strong electron-correlation 
effects including transition metal compounds and organic conductors (for a review, see ref.\onlinecite{S_ImadaMiyake}). 
In the present case, the target bands consist mainly of the Fe $3d$ five orbitals. Detailed derivation of thus obtained 3D low-energy effective models 
is given in refs.\onlinecite{S_Miyake} and \onlinecite{S_Misawa}.
We further downfold the 3D model into the 2D model by computing the screening effect from the layers other than the target layer\cite{S_Nakamura_2D,S_Misawa}.
The validity of our effective model in 2D is confirmed by the fact that 
the magnetic ordered moment is consistent with the experimental values 
as shown in Fig.~1.

Our 2D low-energy model 
is written for ten-fold degenerate Fe-3$d$ 
orbitals in a unit cell containing two Fe atoms in the form
\begin{eqnarray}
\mathcal{H}
&=& \sum_{\sigma} \sum_{i,j} \sum_{\nu,\mu}  
  t_{i,j,\nu,\mu} 
                   a_{i,\nu,\sigma}^{\dagger} 
                   a_{j,\mu,\sigma}   \nonumber \\
&+& \frac{1}{2} \sum_{\sigma, \sigma'} \sum_{i} \sum_{\nu,\mu} 
  \biggl\{ U_{i,i,\mu,\nu} 
                   a_{i,\nu,\sigma}^{\dagger} 
                   a_{i,\mu,\sigma'}^{ \dagger}
                   a_{i,\mu,\sigma'} 
                   a_{i\nu,\sigma} \nonumber \\ 
&+& J_{i,i,\mu,\nu } 
\bigl(\!a_{i,\nu,\sigma}^{\dagger} 
      \!a_{i,\mu,\sigma'}^{ \dagger}
      \!a_{i,\nu,\sigma'} 
      \!a_{i,\mu,\sigma}  \nonumber \\
   \!&+&\!a_{i,\nu,\sigma}^{\dagger} 
      \!a_{i,\nu,\sigma'}^{\dagger}
      \!a_{i,\mu,\sigma'} 
      \!a_{i,\mu,\sigma}\bigr)\! \biggr\},
\label{Eq:Ham}
\end{eqnarray}
where $a_{i,\nu,\sigma}^{\dagger}$ ($a_{i,\nu,\sigma}$) 
is a creation (annihilation) operator of an electron with 
spin $\sigma$ on the $\nu$th maximally localized Wannier orbital~\cite{S_Souza} 
at the $i$-th site. 
$t_{i,j,\nu,\mu}$ contains single-particle levels and transfer integrals.
$U_{i,i,\nu,\mu}$ and 
$J_{i,i,\nu,\mu}$ are screened Coulomb and exchange 
interactions, respectively.  
We use the transfer integrals up to the fifth neighbors, which well reproduce the
LDA band structures.
Offsite interactions are ignored since those are less than a quarter of the onsite parameters and 
we take the site-independent notation $U_{\nu,\mu}$ and $J_{\nu,\mu}$ instead of $U_{i,i,\nu,\mu}$ and $J_{i,i,\nu,\mu}$, respectively hereafter.  
Our results of the $ab$ $initio$ dimensional downfolding for 
each compound all indicate that 
the screenings from the other layers reduce all the Coulomb interactions
uniformly~\cite{S_Nakamura_2D}. Based on our calculated ab initio results, we subtract the constant values from
the $U_{\nu,\mu}$ in 3D models~\cite{S_Miyake} as follows;  
0.44 eV for LaFePO, 0.41 eV for LeFeAsO, 0.38 eV for BaFe$_{2}$As$_{2}$, and
0.40 eV for FeTe, respectively.
We note that the exchange interactions $J_{\nu,\mu}$ are not 
significantly changed by the screenings from the other layers~\cite{S_Nakamura_2D}.
All the interactions parameters that we used in this paper are shown in Table S I.

The double counting of the electron correlation already considered in the local 
density approximation (LDA) is carefully removed with adjusting of the orbital levels so that the 
Hartree solution of effective models conforms with the LDA solution for real materials~\cite{S_Misawa}. 
By using this method, we estimate the double counting part proportional to the Coulomb interactions and
confirm that the inter-orbital part is much smaller than the intra-orbital part.
Therefore neglecting the inter-orbital part hardly changes the present results. 
Furthermore, by following the standard way for estimating the double counting~\cite{S_Anisimov}, 
it is confirmed that the double counting of the exchange correlation is also negligible~\cite{S_Misawa}.
To examine the doping effects, we change the number of electrons without
changing the band structures, i.e., we maintain the orbital levels at those of the undoped ($d^{6}$) case.

\begin{table*} 
\caption{
Effective on-site Coulomb ($U$)/exchange ($J$) interactions between two electrons 
on the same iron site in the $ab$ $initio$ model (in eV). 
}
\ 
\label{tab:Parameters} 
{\scriptsize 
\begin{tabular}{ccccccccccccccc} 
\hline  \\   
LaFePO &      &      & $U$   &      &          &  &     &     &      &  $J$   &      &    \\ 
\hline \\
 & $XY$ & $YZ$ & $Z^2$ & $ZX$ & $X^2-Y^2$  &  &  & $XY$ & $YZ$ & $Z^2$ & $ZX$ & $X^2-Y^2$  \\ 
\hline 
$XY$ & 2.54 & 1.36 & 1.34 & 1.36 & 1.33  &   & $XY$ & & 0.45 & 0.54 & 0.45 & 0.20 \\ 
$YZ$ & 1.36 & 1.98 & 1.53 & 1.20 & 1.02  & & $YZ$ & 0.45 & & 0.32 & 0.36 & 0.31 \\ 
$Z^2$ & 1.34 & 1.53 & 2.37 & 1.53 & 1.02 & & $Z^2$ & 0.54 & 0.32 & & 0.32 & 0.37 \\ 
$ZX$ & 1.36 & 1.20 & 1.53 & 1.98 & 1.02  & & $ZX$ & 0.45 & 0.36 & 0.32 & & 0.31 \\ 
$X^2-Y^2$ & 1.33 & 1.02& 1.02 & 1.02 & 1.24 & & $X^2-Y^2$ & 0.20 & 0.31 & 0.37 & 0.31 & \\ 
\hline  \\ 
LaFeAsO  &      &      & $U$   &      &          &  &     &     &      &  $J$   &      &    \\
\hline \\
        & $XY$ & $YZ$ & $Z^2$ & $ZX$ & $X^2-Y^2$ & &     & $XY$ & $YZ$ & $Z^2$ & $ZX$ & $X^2-Y^2$  \\ 
\hline 
$XY$ & 2.62  & 1.39  & 1.37 & 1.39 & 1.50 &   & $XY$     &      & 0.46 & 0.57 & 0.46 & 0.23 \\ 
$YZ$ & 1.39  & 2.02 & 1.56 & 1.21 & 1.11  &   & $YZ$     & 0.46 &      & 0.33 & 0.37 & 0.35 \\ 
$Z^2$ & 1.37  & 1.56 & 2.43 & 1.56 & 1.10 &   & $Z^2$    & 0.57 & 0.33 &      & 0.33 & 0.42 \\ 
$ZX$ & 1.39  & 1.21 & 1.56 & 2.02 & 1.11  &   & $ZX$     & 0.46 & 0.37 & 0.33 &      & 0.35 \\ 
$X^2-Y^2$ & 1.50  & 1.11 & 1.10 & 1.11 & 1.50 & &$X^2-Y^2$ & 0.23 & 0.35 & 0.42 & 0.35 &      \\
\hline \\
BaFe$_{2}$As$_{2}$  &      &      & $U$   &      &          &  &     &     &      &  $J$   &      &    \\
\hline  \\
 & $XY$ & $YZ$ & $Z^2$ & $ZX$ & $X^2-Y^2$  & & & $XY$ & $YZ$ & $Z^2$ & $ZX$ & $X^2-Y^2$  \\ 
\hline 
$XY$ & 2.80 & 1.56 & 1.61 & 1.56 & 1.78  & & $XY$ & & 0.48 & 0.60 & 0.48 & 0.26 \\ 
$YZ$ & 1.56 & 2.26 & 1.83 & 1.39 & 1.34  & & $YZ$ & 0.48 & & 0.36 & 0.40 & 0.41 \\ 
$Z^2$ & 1.61 & 2.83 & 2.90 & 1.83 & 1.39 & & $Z^2$ & 0.60 & 0.36 & & 0.36 & 0.50 \\ 
$ZX$ & 1.56 & 1.39 & 1.83 & 2.26 & 1.34  & & $ZX$ & 0.48 & 0.40 & 0.36 & & 0.41 \\ 
$X^2-Y^2$ & 1.78 & 1.34 & 1.39 & 1.34 & 1.91  & & $X^2-Y^2$ & 0.26 & 0.41 & 0.50 & 0.41 & \\ 
\hline \\
FeTe &      &      & $U$   &      &          &  &     &     &      &  $J$   &      &    \\ 
\hline  \\
 & $XY$ & $YZ$ & $Z^2$ & $ZX$ & $X^2-Y^2$   &   & & $XY$ & $YZ$ & $Z^2$ & $ZX$ & $X^2-Y^2$   \\
\hline 
$XY$ & 3.44 & 1.94 & 2.10 & 1.94 & 2.64 & & $XY$ & & 0.49 & 0.68 & 0.49 & 0.34 \\ 
$YZ$ & 1.94 & 2.48 & 2.16 & 1.63 & 1.89 & & $YZ$ & 0.49 & & 0.37 & 0.37 & 0.49 \\ 
$Z^2$ & 2.10 & 2.16 & 3.44 & 2.17 & 2.04 & & $Z^2$ & 0.68 & 0.37 & & 0.37 & 0.66 \\ 
$ZX$ & 1.94 & 1.63 & 2.17 & 2.48 & 1.89 & & $ZX$ & 0.49 & 0.37 & 0.37 & & 0.49 \\ 
$X^2-Y^2$ & 2.64 & 1.89 & 2.04 & 1.89 & 3.19 & & $X^2-Y^2$ & 0.34 & 0.49 & 0.66 & 0.49 & \\ 
\hline 
\end{tabular} 
} 
\end{table*}

\section{S.2 Discussion on adequacy of single parameter scaling by $\lambda$}
 
Although the effective Coulomb interaction $U_{\nu,\mu}$ rather strongly depends on combinations of electronic orbitals $\nu,\mu$, 
the ratio to the $3d$ bandwidth has a clear systematic and uniform change from compounds to compounds with relatively small variance.  
We introduce the dominant transfer by $\bar{t} =\frac{1}{5}\sum_{\nu=1,5}t_{{\rm max},\nu}$, where $t_{{\rm max,\nu}}$ is the largest transfer among pairs between 
an orbital $\nu$ at an iron site and any 3$d$ orbitals at its nearest-neighbor iron site.  Then 
the scaled ratio of the interaction strength is defined by the dimensionless parameter $\bar{U}_{\nu\mu} =U_{\nu\mu}/\bar{t}$.  
A single parameter to characterize the ratio of the interaction to 
the bandwidth normalized by the value for LaFeAsO is 
\begin{align}
\lambda&=\frac{\sum_{\nu,\mu}\bar{U}_{\nu\mu}}{\sum_{\nu,\mu}\bar{U}_{{\rm LFAO}\nu\mu}}, 
\label{lambda}
\end{align}
where $\bar{U}_{{\rm LFAO}\nu\mu}$ in the denominator is the orbital dependent $\bar{U}_{\nu\mu}$ for LaFeAsO. 
Then $\lambda$ monitors the orbital averaged ``$U/t$" scaled by the value for LaFeAsO. 
We note that the values of $\lambda$ for each materials are
slightly different from those of ref.~\onlinecite{S_Misawa}, because
In ref.~\onlinecite{S_Misawa}, we have performed
the dimensional downfolding only for LaFeAsO and have assumed
the reduction of Coulomb interactions is independent of materials.
In this paper, we have performed the
dimensional downfolding for each material, not only for LaFeAsO,
but also for LaFePO, BaFe$_2$As$_2$ and FeTe to derive the
{\it ab initio} two-dimensional models, and we have precisely
determined the reductions of Coulomb interactions depending on
materials. The reductions slightly depend on materials as one sees
in Section S.1. Small material dependences of the reduction
(within 10\% of the whole reduction) is the reason why the estimation of $\lambda$ is slightly 
different from that of ref.~\onlinecite{S_Misawa}.
We note that this slight difference did not essentially alter out results. 

Since the band structure of the Fe 3$d$ bands near the Fermi level is similar for all the 
iron-based superconductors with a similar bandwidth around 4.5 eV, the ratio $\lambda$ is 
the only prominent parameter to identify the material character of each compound and 
it is indeed a measure for the electron correlation strength. 

To see whether the single parameter $\lambda$ well represents the material dependence, we introduce quantities
$I_{U\nu\mu}\equiv\bar{U}_{\nu\mu}/\bar{U}_{{\rm LFAO}\nu\mu}$,
$I_U\equiv\frac{1}{25}\sum_{\nu,\mu}\bar{U}_{\nu\mu}/\bar{U}_{{\rm LFAO}\nu\mu}$
$I_{J\nu\mu}\equiv\bar{J}_{\nu\mu}/\bar{J}_{{\rm LFAO}\nu\mu}$,
and
$I_J\equiv\frac{1}{25}\sum_{\nu,\mu}\bar{J}_{\nu\mu}/\bar{J}_{{\rm LFAO}\nu\mu}$
with
$\bar{J}_{\nu\mu} =J_{\nu\mu}/\bar{t}$.
If $I_U/\lambda$ and $I_J/\lambda$ are close to unity, it supports that 
the interactions of each material is well described by scaling the interactions of the
reference LaFeAsO.
Further support can be obtained by examining the variances 
\begin{align}
\Delta I_U&=\sqrt{\frac{1}{25}\sum_{\nu,\mu}[I_{U\nu\mu}-I_U]^2}, \\ 
\Delta I_J&=\sqrt{\frac{1}{25}\sum_{\nu,\mu}[I_{J\nu\mu}-I_J]^2}.
\label{lambda}
\end{align}
In Fig.~S\ref{fig:ratio}, $I_U$, and $I_J$ for several different materials at the corresponding $\lambda$ indeed show that 
$I_U/\lambda$ and $I_J/\lambda$ are close to unity for all the materials.
This means that interaction parameters are well scaled uniformly with 
the single parameter $\lambda$ for different families.
A relatively large deviation from $I=\lambda$ with a substantial variance is seen for $I_J$ of FeTe. However, the magnetic moment of FeTe is more or 
less already saturated because of  the large $U$, and this variance does not seriously influence the prediction on the moment.
\renewcommand{\figurename}{Fig. S}
\setcounter{figure}{0}
\begin{figure}[h!]
     \begin{center}
       \includegraphics[width=7cm,clip]{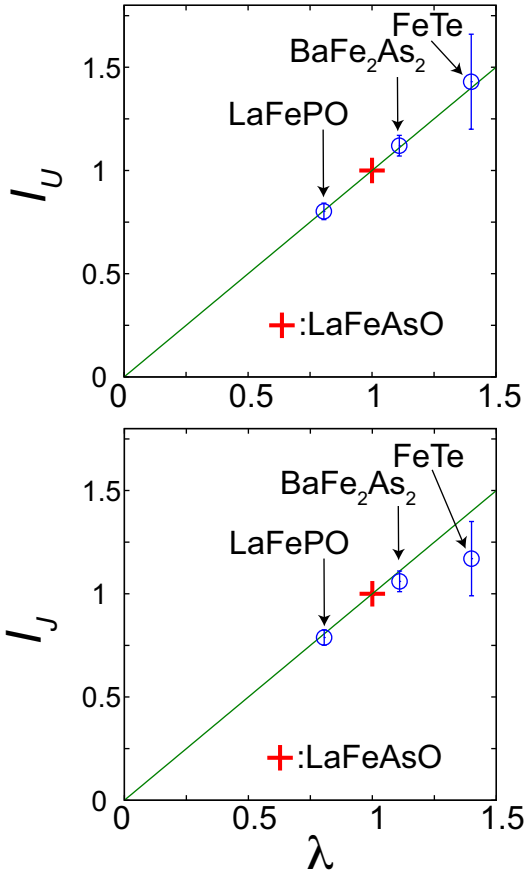}
     \end{center}
     \caption{ Scaling plot of interaction parameters for several different families. 
 Definition of $I$ and $\lambda$ are given in the text and the error bars show $\Delta I_U$ and $\Delta I_J$, 
the variance (standard deviation) of $I_U$ and $I_J$, respectively. The single parameter 
scaling by $\lambda$ works well if the data are on the line $I=\lambda$ with small variances. We admit roughly $5\%$ error in the assignments of the real materials.
}
     \label{fig:ratio}
  \end{figure}%

\vskip 5mm
\noindent
\section{S.3 MULTI-VARIABLE VARIATIONAL MONTE CARLO METHOD}

Our variational wave function is defined as
\begin{equation}
|\psi\ra = \sP_{\rm G}\sP_{\rm J}\sL^{S=0}|\phi_{\rm pair}\ra,
\label{Eq:WF}
\end{equation}
where $\sP_{\rm G}$ and $\sP_{\rm J}$ are the Gutzwiller and Jastrow factors, respectively~\cite{S_Tahara}.
The Gutzwiller factor punishes the double occupation of electrons on the same orbital defined by the orbital-dependent variational parameters $g_{\nu}$ as
\begin{align*}
  \sP_{\text{G}} &= \exp \biggl[
    -\sum_{i,\nu} g_{\nu}n_{i\nu\uparrow} n_{i\nu\downarrow}
  \biggr] \\ 
&= \prod_{i,\nu} \Bigl[ 1 - (1-e^{-g_{\nu}}) n_{i\nu\uparrow} n_{i\nu\downarrow} \Bigr]
,
\end{align*}
where $n_{i\nu\sigma}$ is the density operator for the orbital $\nu$ with the spin $\sigma$ at the $i$-th site, and the Jastrow factor takes into account the interorbital and/or intersite 
charge correlations by the site- and orbital-dependent variational parameters $v_{ij\nu\mu}$ as
\begin{equation}
  \sP_{\text{J}} = \exp \biggl[
    - \frac{1}{2} \sum_{i,j,\nu,\mu} v_{ij\nu\mu} n_{i\nu} n_{j\mu}.
  \biggr] 
\end{equation}
In the present study, we take into account up to the nearest neighbor for the intersite pair $(i,j)$ including the onsite interorbital pairs by considering all the combinations of orbital dependent correlations.

\begin{figure}[tb]
     \begin{center}
       \includegraphics[width=7cm,clip]{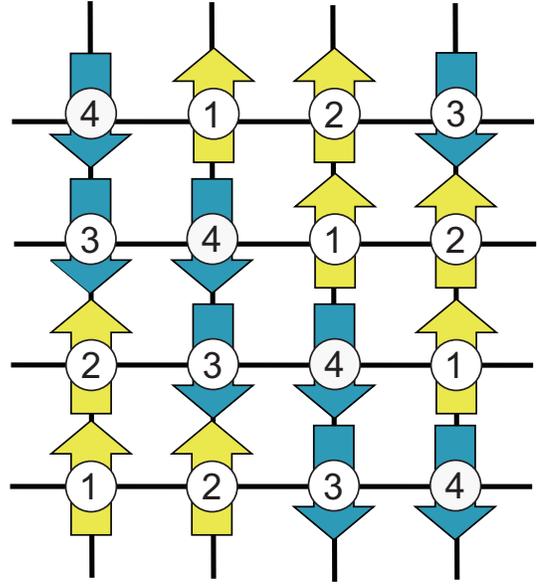}
     \end{center}
     \caption{ 
      The translational symmetry employed for tiling by $1\times4$ supercell structure. 
      The labels 1-4 denote the sublattice indices for the $1\times4$ supercell structure for $f_{ij}$. 
      The arrows indicate the pattern of the staggered magnetic ordered
      moment realized in the present result of the double-stripe structure.} 
     \label{fig:SubLattice}
\end{figure}%

The spin quantum-number projection $\sL^{S=0}$ restores the
SU(2) spin-rotational symmetry and generates a state with the
correct total spin $S$=$0$.
The one-body part $|\phi_{\rm pair}\ra$ is
the generalized pairing wave function defined as
$|\phi_{\rm pair}\ra=\Big[\sum_{i,j=1}^{\Ns}f_{ij}c_{i\uparrow}^{\dag}c_{j\downarrow}^{\dag}\Big]^{N/2} |0 \ra$
with $f_{ij}$ being the variational parameters. Here, $N_s$ and $N$ are the system size (number of sites) and the number of electrons, respectively.   
In this study, we 
allow $f_{ij}$ to have $2\times2$ supercell structure
or equivalently we have $2\times2\times5^{2}\times N_{\rm s}$ independent variational parameters.
All the variational parameters are simultaneously 
optimized by using the stochastic reconfiguration method~\cite{S_Tahara,S_Sorella}.
The variational function $|\psi\ra$ 
in Eq.(\ref{Eq:WF}) 
can flexibly describe paramagnetic metals and the stripe type AF and G-type AF phases, as well as
superconducting phases and correlated metals. In the present calculation we also examined the 
possibility of the double stripe type order realized for FeTe by modifying $f_{i,j}$ with the $1\times 4$ 
supercell (see Fig.~S \ref{fig:SubLattice}), 
while we have restricted to the normal state for the metals. For the accuracy of the present method see ref.\onlinecite{S_Tahara}.

The magnetic ordered moment is defined by
$m(\bm{q})^{2}\equiv {\frac{4}{3N_{\rm s}^2} \sum_{i,j}\langle\bm{S}_{i}\bm{S}_{j}\rangle e^{i\bm{q(r_{i}-r_{j})}}}$
for $N_{\rm s}$-site ($N_{\rm s}/2$ unit cell) system with the periodic 
boundary condition and $\bm{q}$ is the momentum at the peak, which is $q_{\rm peak}\equiv(0,\pi)$ and $(\pi,\pi)$ and $(\pi/2,\pi/2)$ for the stripe, G-type and double-stripe orders, respectively.
In this definition, the saturated magnetic moment for the classical N\'{e}el state is
given by $m$ = 4 $\mu_{\rm B}$.
The double occupation is defined by 
$D_{\nu}\equiv {\frac{1}{N_{\rm s}} \sum_{i}\langle n_{i,\nu,\uparrow}n_{i,\nu,\downarrow}\rangle}$
and the orbital dependent momentum distribution $n_{\nu}(k)$ and orbital-dependent filling $n_{\nu}$ are defined by
$n_{\nu}\equiv \frac{1}{N_{\rm s}} \sum_{k}n_{\nu}(k)$ and 
$n_{\nu}(k)\equiv  \sum_\sigma \langle a^{\dagger}_{\nu,\sigma}(k)a_{\nu,\sigma}(k)\rangle$.
All the physical quantities are presented after proper size extrapolations to the thermodynamic limit unless explicitly stated otherwise.

\end{document}